\magnification=1200  	
\nopagenumbers
\pageno=-1
\baselineskip=18pt plus 1pt minus 1pt
\parindent=23pt
\font\csc=cmcsc10
\font\eightrm=cmr8 at 8pt


\rightline{\eightrm February 1996}
\vskip57pt
\centerline{\csc\bf Ricci Tensor of Diagonal Metric}
\vskip27pt
\centerline{K.~Z.~Win}
\centerline{\it Department of Physics and Astronomy}
\centerline{\it University of Massachusetts}
\centerline{\it Amherst, MA 01003}

\vskip87pt
\centerline{\bf Abstract}

\noindent
\centerline {Efficient formulae of Ricci tensor for an arbitrary diagonal 
metric are presented.}
\vfill
\eject

\pageno=1\footline={\rm\hfil\folio\hfil}

\noindent
{\bf Introduction}

Calulation of the Ricci tensor is often a cumbersome task.  In 
this note useful formulae of the Ricci tensor are presented in equations (1) 
and (2) 
for the case of the diagonal metric tensor.   Application of the formulae in 
computing the Ricci tensor of $n$-sphere is also presented.

\vskip9pt
\noindent
{\bf Derivation}
 
The sign conventions and notation of Wald [1] will be followed.
Latin indices  are part of abstract index notation and Greek 
indices denote basis components.  The Riemann curvature tensor
$R_{abc}\thinspace^d$  is 
defined by $ \nabla_a\nabla_b\omega_c - \nabla_b\nabla_a\omega_c 
=R_{abc}\thinspace^d\omega_d $.  The Ricci tensor 
in a coordinate basis is 
$R_{\lambda\nu}=\partial_\rho\Gamma^\rho_{\lambda\nu} - 
\partial_\lambda\Gamma^\rho_{\nu\rho} 
- \Gamma_{\lambda\rho}^\sigma\Gamma^\rho_{\sigma\nu}
+ \Gamma_{\lambda\nu}^\rho\Gamma_{\rho\sigma}^\sigma $
where the  Christoffel symbols are
$\Gamma^\rho_{\lambda\nu}= g^{\rho\sigma}\left(
\partial_\lambda g_{\nu\sigma} +
\partial_\nu g_{\lambda\sigma} -
\partial_\sigma g_{\lambda\nu}\right)/2 $.  
{\it For the rest of this note we assume that $g_{\rho\nu}$ is diagonal 
unless otherwise indicated.}
Also to avoid having to write down ``$\mu\neq\nu$'' repeatedly we will 
{\it always} take  $\mu\neq\nu$  and no sum will be 
assumed on repeated  $\mu$ and $\nu$.

We first note that $\Gamma^{\rho}_{\mu\nu}=0$ if 
$\mu\neq\rho\neq\nu$.  It means that at least two indices of 
$\Gamma$ must be the same for it to be nonvanishing.  Also $\ln |g| =
\sum_{\sigma=1}^n \ln|g_{\sigma\sigma}|$ where $g$ and $n$ are  
determinant of the 
metric and dimension of manifold, respectively. 
  Then it is easy to show that $2\Gamma_{\sigma\mu}^\mu = 
\partial_\sigma \ln|g_{\mu\mu}|$ for all $\sigma$ and that 
$2\Gamma_{\mu\mu}^{\nu}=-g^{\nu\nu}\partial_\nu g_{\mu\mu}$.
(Again $\mu\neq\nu$ and no sum on $\mu$ and $\nu$.)  These two equations 
give 

$
4\Gamma_{\mu\mu}^{\nu}\Gamma_{\nu\nu}^{\mu} = 
g^{\nu\nu}\left(\partial_\nu g_{\mu\mu}\right)
g^{\mu\mu}\left(\partial_\mu g_{\nu\nu}\right)
=\left(\partial_\mu\ln|g_{\nu\nu}|\right)
\left(\partial_\mu\ln|g_{\nu\nu}|\right)
=4\Gamma^\mu_{\mu\nu}\Gamma^\nu_{\mu\nu}$

\noindent
Using above facts, we  get, with each sum having upper limit 
$n$ and lower limit 1, $$\eqalign {
R_{\mu\nu}  =& 
\partial_\mu\Gamma^{\mu}_{\mu\nu} + \partial_\nu\Gamma^\nu_{\mu\nu}
 - (1/2)\partial_\mu\partial_\nu\ln|g| - 
\Gamma^\mu_{\mu\rho}\Gamma^\rho_{\mu\nu} -
\Gamma^\nu_{\mu\rho}\Gamma^\rho_{\nu\nu} - 
\sum_{\sigma\neq\mu,\nu} 
\Gamma^\sigma_{\mu\rho}\Gamma^\rho_{\sigma\nu} \cr
&\qquad\quad+(1/ 2)\Gamma^\mu_{\mu\nu}\partial_\mu\ln|g| +
(1/ 2)\Gamma^\nu_{\mu\nu}\partial_\nu\ln|g| \cr
=&(1/ 2)\partial_\mu\partial_\nu \ln |{g_{\mu\mu}g_{\nu\nu}\over 
g}| 
- \Gamma^\mu_{\mu\mu}\Gamma^{\mu}_{\mu\nu} 
- \Gamma^\mu_{\mu\nu}\Gamma^{\nu}_{\mu\nu} 
- \Gamma^\nu_{\mu\mu}\Gamma^{\mu}_{\nu\nu} 
- \Gamma^\nu_{\nu\mu}\Gamma^{\nu}_{\nu\nu}
-\sum_{\sigma\neq\mu,\nu}  
\Gamma^\sigma_{\sigma\mu}\Gamma^{\sigma}_{\sigma\nu} \cr
&\quad+ \Gamma^\mu_{\mu\nu}\partial_\mu\ln\sqrt{|g_{\mu\mu}|}
+ \Gamma^\mu_{\mu\nu}\partial_\mu\ln\sqrt{|g_{\nu\nu}|}
+  \Gamma^\nu_{\mu\nu}\partial_\nu\ln\sqrt{|g_{\mu\mu}|}
+ \Gamma^\nu_{\mu\nu}\partial_\nu\ln\sqrt{|g_{\nu\nu}|}\cr
&\quad+(1/2) 
\Gamma^\mu_{\mu\nu}\partial_\mu\ln|{g\over 
g_{\mu\mu}g_{\nu\nu}}|
+(1/2)\Gamma^\nu_{\mu\nu}\partial_\nu\ln|{g\over g_{\mu\mu}g_{\nu\nu}}| 
\cr}$$
\vfil
\eject
$$\eqalign{
=&(1/2)\partial_\mu\partial_\nu \ln |{g_{\mu\mu}g_{\nu\nu}\over 
g}| 
- \Gamma^\mu_{\mu\mu}\Gamma^{\mu}_{\mu\nu}
- \Gamma^\mu_{\mu\nu}\Gamma^{\nu}_{\mu\nu} 
- \Gamma^\nu_{\mu\nu}\Gamma^{\mu}_{\mu\nu} 
- \Gamma^\nu_{\nu\mu}\Gamma^{\nu}_{\nu\nu}-
\sum_{\sigma\neq\mu,\nu}  
\Gamma^\sigma_{\sigma\mu}\Gamma^{\sigma}_{\sigma\nu} \cr
&+ \Gamma^\mu_{\mu\mu}\Gamma^{\mu}_{\mu\nu} 
+ \Gamma^\mu_{\mu\nu}\Gamma^{\nu}_{\mu\nu} 
+ \Gamma^\nu_{\mu\nu}\Gamma^{\mu}_{\mu\nu}
+ \Gamma^\nu_{\nu\mu}\Gamma^{\nu}_{\nu\nu}\cr
&+(1/2)
\Gamma^\mu_{\mu\nu}\partial_\mu\ln|{g\over g_{\mu\mu}g_{\nu\nu}}|
+(1/2)\Gamma^\nu_{\mu\nu}\partial_\nu\ln|{g\over 
g_{\mu\mu}g_{\nu\nu}}| \cr
 =&(1/2)\partial_\mu\partial_\nu \ln |{g_{\mu\mu}g_{\nu\nu}\over 
g}|  -
\sum_{\sigma\neq\mu,\nu}  
\Gamma^\sigma_{\sigma\mu}\Gamma^{\sigma}_{\sigma\nu} 
+{1\over 2}\Gamma^\mu_{\mu\nu}\partial_\mu\ln|{g\over 
g_{\mu\mu}g_{\nu\nu}}|
+{1\over 2}\Gamma^\nu_{\mu\nu}\partial_\nu\ln|{g\over 
g_{\mu\mu}g_{\nu\nu}}|\cr}$$
or (remember $\mu\neq\nu$)
$$
4R_{\mu\nu} =
\left(\partial_\mu\ln|g_{\nu\nu}|-\partial_\mu\right)
\partial_\nu\ln|{g\over g_{\mu\mu}g_{\nu\nu}}|
+(\mu\leftrightarrow\nu)
-\sum_{\sigma\neq\mu,\nu}
\partial_\mu\ln|g_{\sigma\sigma}|
\partial_\nu\ln|g_{\sigma\sigma}| \eqno(1)
$$
where $(\mu\leftrightarrow\nu)$ stands for preceding terms with $\mu$ and 
$\nu$ interchanged.\footnote*{Alternative forms of equation (1) are

$4R_{\mu\nu} = \sum_{\sigma\neq\mu,\nu}
\left[\left(\partial_\mu\ln|g_{\nu\nu}| - \partial_\mu\right)
\partial_\nu \ln|g_{\sigma\sigma}|
+(\mu\leftrightarrow\nu)-
\partial_\mu\ln|g_{\sigma\sigma}|
\partial_\nu\ln|g_{\sigma\sigma}|\right]
$
and

$8R_{\mu\nu}=\sum_{\sigma\neq\mu,\nu}\left[
\left(\partial_\mu\ln{g^2_{\nu\nu}\over |g_{\sigma\sigma}|}-2\partial_\mu
\right)\partial_\nu\ln|g_{\sigma\sigma}|+(\mu\leftrightarrow\nu)\right]$}
  A 
number of useful facts follow 
from the above formula.  
  First of all {\it $R_{\mu\nu}=0$ if $x_\mu$ is an ignorable coordinate} 
because $\partial_\mu$ is in each term of the formula.  If there are 
$m$ ignorable coordinates then maximum number of nonzero off-diagonal
components of the 
Ricci tensor is $(n-m)(n-m-1)/2$.
Thus a sufficient condition for the Ricci
 tensor to be diagonal is that  the (diagonal) 
metric be a function  only of one 
coordinate.
The Robertson-Walker metric with flat spatial sections,
 $ds^2=-dt^2+a(t)^2(dx^2+dy^2+dz^2)$, satisfies this condition and its Ricci 
tensor is consequently diagonal. If the (diagonal)
metric 
is a 
function only of two coordinates, say  $x_\mu$ and $x_\nu$, then all off 
diagonal components of the Ricci tensor except $R_{\mu\nu}$ are zero.  For 
example,
for Schwartzschild metric $ds^2=-B(r)dt^2 
+A(r)dr^2+r^2\left(d\theta^2+\sin^2\theta d\phi^2\right)$ only $R_{\theta 
r}$ needs to be explicitly calculated, all other being zero.  It is 
easily checked using equation (1) that $R_{\theta r}$ is also zero.   
Symmetry 
arguments can also be given\footnote\dag{Page 178 of Weinberg [2].}
 as to 
why the Ricci tensor is diagonal in 
this case.
Secondly {\it second derivative terms are 
identically zero whenever each  component of the metric
  is a product of functions of one single coordinate} i.e.~if, 
for all $\mu$, $g_{\mu\mu}=\prod_{l=1}^{n}f_{\mu}^l
(x_l)$ where 
some of 
$f^l_\mu$ may be one. 
A final trivial corollary of equation (1)  is that  
the Ricci tensor is 
 diagonal  in  2-dimensions.  This  fact also follows trivially 
from 
the fact that in 2-dimensions, the Ricci tensor is the metric tensor (not 
necessarily diagonal) up to a factor of a
scalar function.  When $n=3$, it is easy to find an example of a diagonal 
 metric which results in a non-diagonal Ricci tensor. For example,
 $ds^2 = dx^2 +xdy^2+ydz^2$ gives $4R_{xy}=1/xy$.  

We now calculate the diagonal components of Ricci tensor.  We have
$$\eqalign{
R_{\mu\mu}=&
\partial_\sigma\Gamma_{\mu\mu}^\sigma-
\partial_\mu^2\ln\sqrt{|g|}
-\Gamma_{\mu\rho}^\sigma\Gamma^\rho_{\sigma\mu}
+\Gamma_{\mu\mu}^\rho\partial_\rho\ln\sqrt{|g|}\cr
=&\partial_\mu^2\ln\sqrt{|g_{\mu\mu}|}
+\sum_{\sigma\neq\mu}\partial_\sigma\Gamma^\sigma_{\mu\mu}
-\partial_\mu^2\ln\sqrt{|g|}
-\Gamma_{\mu\rho}^\mu\Gamma^\rho_{\mu\mu}\cr
&\qquad-\sum_{\sigma\neq\mu}\Gamma^\sigma_{\mu\rho}\Gamma^\rho_{\mu\sigma}
+\Gamma^\mu_{\mu\mu}\partial_\mu\ln\sqrt{|g|}
+\sum_{\sigma\neq\mu}\Gamma_{\mu\mu}^\sigma\partial_\sigma \ln\sqrt{|g|}\cr
=&(1/2)\partial_\mu^2\ln|{g_{\mu\mu}\over g}|
+\sum_{\sigma\neq\mu} \partial_\sigma\Gamma^\sigma_{\mu\mu} 
-\left(\Gamma^\mu_{\mu\mu}\right)^2 
- \sum_{\sigma\neq\mu}\Gamma^\mu_{\mu\sigma}\Gamma^\sigma_{\mu\mu}
- \sum_{\sigma\neq\mu}\Gamma^\sigma_{\mu\mu}\Gamma^\mu_{\sigma\mu}\cr
&\quad-\sum_{\sigma\neq\mu}\Gamma^\sigma_{\sigma\mu}\Gamma^\sigma_{\sigma\mu}
+\Gamma_{\mu\mu}^\mu\partial_\mu\ln\sqrt{|g_{\mu\mu}|}
+\Gamma_{\mu\mu}^\mu\partial_\mu\ln\sqrt{|{g\over g_{\mu\mu}}|}
+\sum_{\sigma\neq\mu} \Gamma^\sigma_{\mu\mu}\partial_\sigma\ln\sqrt{|g|} \cr
=&(1/4)\left(\partial_\mu\ln|g_{\mu\mu}|
-2\partial_\mu\right)\partial_\mu\ln|{g\over g_{\mu\mu}}| 
+\sum_{\sigma\neq\mu}\Big[\partial_\sigma\Gamma^\sigma_{\mu\mu} 
- 2\Gamma_{\mu\mu}^\sigma\Gamma^\mu_{\mu\sigma}
-\left(\Gamma^\sigma_{\sigma\mu}\right)^2\cr
&\qquad
\qquad\qquad\qquad
\qquad\qquad+\Gamma^\sigma_{\mu\mu}\partial_\sigma\ln\sqrt{|g|}\Big] \cr
=&(1/4)\left(\partial_\mu\ln|g_{\mu\mu}|
-2\partial_\mu\right)\partial_\mu\ln|{g\over g_{\mu\mu}}| 
+\sum_{\sigma\neq\mu}\bigg[
-(1/2)\partial_\sigma\left(g^{\sigma\sigma}\partial_\sigma 
g_{\mu\mu}\right)\cr
&\quad+g^{\sigma\sigma}\left(\partial_\sigma 
g_{\mu\mu}\right)\partial_\sigma\ln\sqrt{| g_{\mu\mu}|}
-\left(\partial_\mu\ln\sqrt{|g_{\sigma\sigma}|}\right)^2
-(g^{\sigma\sigma}/ 2)(\partial_\sigma g_{\mu\mu})\partial_\sigma
\ln\sqrt{|g|}\bigg]\cr
=&{1\over 4}\left(\partial_\mu\ln|g_{\mu\mu}|-2\partial_\mu\right)
\partial_\mu\ln|{g \over g_{\mu\mu}}|
-{1\over 4} \sum_{\sigma\neq\mu}\bigg[ 2\partial_\sigma
\left(g^{\sigma\sigma}\partial_\sigma g_{\mu\mu}\right)
+\left(\partial_\mu\ln|g_{\sigma\sigma}|\right)^2         \cr
&\qquad\qquad
+g^{\sigma\sigma}\left(\partial_\sigma g_{\mu\mu}\right)   
\partial_\sigma\ln {|g|\over g^2_{\mu\mu}}\bigg]         \cr
}$$
or
$$
4R_{\mu\mu}= \left(\partial_\mu\ln|g_{\mu\mu}|- 
 2\partial_\mu\right)\partial_\mu\ln|{ g \over g_{\mu\mu} }|
-\sum_{\sigma\neq\mu}\Big[
\left(\partial_\mu\ln|g_{\sigma\sigma}|\right)^2
+\Big(\partial_\sigma\ln{|g|\over g_{\mu\mu}^2}+2\partial_\sigma\Big)
g^{\sigma\sigma}\partial_\sigma g_{\mu\mu}
\Big]  \eqno(2)
$$
where $2\partial_\sigma$ acts everything on its right.  The efficiency of 
this formula is obvious for Schwartzschild metric given earlier.   
To compute $R_{tt}$ we note 
 that all terms with $\partial_t$ are zero.  Thus $
4R_{tt}=-\sum_{\sigma\neq t}\left(\partial_\sigma\ln{|g|\over
 g_{tt}^2} + 2 \partial_\sigma\right)
\left(g^{\sigma\sigma}\partial_\sigma g_{tt}\right)$.
Because $g_{tt}=-B(r)$ only the $\sigma=r$ term survives and we get, 
after very little computation, $
R_{tt}={B'\over Ar} -{B'\over 4A}
\left({A'\over A}+ {B'\over B}\right)
 + {B''\over 2A}   $ where prime indicates
 differentiation with respect 
to $r$.\footnote\ddag{Cf.
page 178 of Weinberg and note that his  $R_{abc}\thinspace^d$ is 
minus of Wald's.  Also see problem 6.2 of [1].}  Other diagonal
components 
are just as easy to compute. 
\vskip9pt
\noindent
{\bf Application to $n$-sphere}

To illustrate the use of above formulae we compute Ricci tensor of 
$n$-sphere which in spherical coordinates has diagonal metric 
 $g_{11}=1$ and 
$g_{\mu+1,\mu+1}=g_{\mu\mu}\sin^2x_\mu$ for $1\leq\mu\leq n-1$.
  First note that 
 $\ln|g_{\mu\mu}|=\sum_{\nu=1}^{\mu-1}2\ln|\sin x_\nu|$
 and thus
 $\partial_\rho g_{\mu\mu} = 0$ if $n\geq\rho\geq\mu\geq 1$.  
These 
facts will be used repeatedly in what follows. We first compute  the off 
diagonal 
components of Ricci tensor.  According to the general 
argument given earlier second derivative terms will be zero.  Let  
$\mu>\nu$.  Then  equation 
(1) becomes 
$$\eqalign{
 4R_{\mu\nu} 
=&\sum_{\sigma>\mu}\left[ 
\left(\partial_\nu\ln|g_{\mu\mu}|\right)\partial_\mu\ln|g_{\sigma\sigma}|
-\left(\partial_\mu\ln|g_{\sigma\sigma}|\right)\partial_\nu\ln
|g_{\sigma\sigma}|\right]\cr
=& 4\sum_{\sigma>\mu}
\left(\cot x_\nu \cot x_\mu - \cot x_\mu\cot x_\nu\right) = 0\cr}$$
	We next compute the diagonal components.  We have, from equation (2), 
$$\eqalign{
4R_{\mu\mu}
=&-\sum_{\sigma>\mu}\Big[
2\partial_\mu^2\ln{|g_{\sigma\sigma}|}
+\left(\partial_\mu\ln|g_{\sigma\sigma}|\right)^2\Big]
-\sum_{\sigma<\mu}
\left(\partial_\sigma\ln{|g|\over g^2_{\mu\mu}}+2\partial_\sigma\right)
g^{\sigma\sigma}\partial_\sigma g_{\mu\mu}\cr
=&\sum_{\sigma>\mu}
\left(-2\times 2\partial_\mu\cot x_\mu - 4\cot^2x_\mu\right)
-\bigg(\partial_{\mu-1}\ln{|g|\over g^2_{\mu\mu}}
+2\partial_{\mu-1}\bigg)
g^{\mu-1,\mu-1}\partial_{\mu-1}g_{\mu\mu}\cr
&\quad-\sum_{\sigma<\mu-1}
\bigg(\partial_\sigma\ln{|g|\over g^2_{\mu-1,\mu-1}}
-4\partial_\sigma\ln|\sin x_{\mu-1}|
+2\partial_\sigma\bigg)
g^{\sigma\sigma}\partial_\sigma
\left(g_{\mu-1,\mu-1}\sin^2 x_{\mu-1}\right)\cr
=&\sum_{\sigma>\mu}
\left({4\over \sin^2x_\mu}-4{\cos^2x_\mu\over \sin^2x_\mu}\right)\cr
&\qquad-\bigg[
\big(\sum_{\sigma>\mu-1}\partial_{\mu-1}\ln|g_{\sigma\sigma}|\big)
-2\partial_{\mu-1}\ln|g_{\mu\mu}|
+2\partial_{\mu-1}\bigg]
g^{\mu-1,\mu-1}\partial_{\mu-1}g_{\mu\mu}\cr
&\qquad-\sin^2x_{\mu-1}\sum_{\sigma<\mu-1}
\big(\partial_\sigma \ln{|g|\over g^2_{\mu-1,\mu-1}}
+2\partial_\sigma\big)
g^{\sigma\sigma}\partial_\sigma g_{\mu-1,\mu-1}\cr
}$$
\vfill\eject
$$\eqalign{
=&4(n-\mu)-\Big[
\big(2 \cot x_{\mu-1}\sum_{\sigma>\mu-1}\big)
-4\cot x_{\mu-1} + 2\partial_{\mu-1}\Big]
g^{\mu-1,\mu-1}g_{\mu-1,\mu-1} \sin2x_{\mu-1}\cr
&\qquad-\sin^2x_{\mu-1}\sum_{\sigma<\mu-1}
\Big(\partial_\sigma\ln{|g|\over g^2_{\mu-1,\mu-1}}
+2\partial_\sigma\Big)
g^{\sigma\sigma}\partial_\sigma g_{\mu-1,\mu-1}\cr
=&4(n-\mu)-\left[2\cot x_{\mu-1}(n+1-\mu)-4\cot x_{\mu-1}\right]
\sin 2x_{\mu-1}\
-2\times 2\cos 2x_{\mu-1}\cr
&\qquad-\sin^2x_{\mu-1}
\sum_{\sigma<\mu-1}
\big(\partial_\sigma\ln{|g|\over 
g^2_{\mu-1,\mu-1}}+2\partial_\sigma\big)
g^{\sigma\sigma}\partial_\sigma g_{\mu-1,\mu-1}\cr
=&4(n-\mu)-4
\big[\cos^2x_{\mu-1}(n-\mu+1)-2\cos^2x_{\mu-1} 
+\cos^2x_{\mu-1}-\sin^2x_{\mu-1}\big]\cr
&\qquad-\sin^2x_{\mu-1}\sum_{\sigma<\mu-1}
\big(\partial_\sigma\ln{|g|\over g^2_{\mu-1,\mu-1}}
+2\partial_\sigma\big)
g^{\sigma\sigma}\partial_\sigma g_{\mu-1,\mu-1}\cr
=&
4\left[n-\mu-\cos^2x_{\mu-1}(n-\mu+1)+1\right]\cr
&\qquad-\sin^2x_{\mu-1}\sum_{\sigma<\mu-1}
\big(\partial_\sigma\ln{|g|\over 
g^2_{\mu-1,\mu-1}}+2\partial_\sigma\big)
g^{\sigma\sigma}\partial_\sigma g_{\mu-1,\mu-1}\cr
=&\sin^2x_{\mu-1}
\Big[4(n-\mu+1)-\sum_{\sigma<\mu-1}
\Big(\partial_\sigma\ln{|g|\over g^2_{\mu-1,\mu-1}}+2\partial_\sigma\Big)
g^{\sigma\sigma}\partial_\sigma g_{\mu-1,\mu-1}\Big]\cr
}
$$
Thus
$$4R_{\mu+1,\mu+1}=
\sin^2x_\mu\bigg[4(n-\mu)-\sum_{\sigma<\mu}
\left(\partial_\sigma\ln{|g|\over g^2_{\mu\mu}} + 2 \partial_\sigma\right)
g^{\sigma\sigma}\partial_\sigma g_{\mu\mu}\bigg]
$$
We now rewrite  the expression inside the square brackets.
 Break up the sum into $\sigma=\mu-1$ and $\sigma<\mu-1$ and 
use $g_{\mu\mu}= g_{\mu-1,\mu-1}\sin^2x_{\mu-1}$.  Then
$$\eqalign{
[\quad]=&4(n-\mu)
-\big(\partial_{\mu-1}\ln{|g|\over g^2_{\mu\mu}}
+2\partial_{\mu-1}\big)
g^{\mu-1,\mu-1}\partial_{\mu-1}g_{\mu\mu}\cr
&\qquad-\sum_{\sigma<\mu-1}
\bigg(\partial_\sigma\ln{|g|\over g^2_{\mu-1,\mu-1}} 
-4\partial_\sigma\ln|\sin x_{\mu-1}|+
2\partial_\sigma\bigg)
g^{\sigma\sigma}\partial_\sigma g_{\mu\mu}\cr
=&4(n-\mu)-\bigg(\partial_{\mu-1}\ln 
|g|-2\partial_{\mu-1}\ln|g_{\mu\mu}| 
+2\partial_{\mu-1}\bigg)g^{\mu-1,\mu-1}g_{\mu-1,\mu-1}\sin 2x_{\mu-1}\cr
&\qquad-\sum_{\sigma<\mu-1}\bigg(\partial_\sigma
\ln{|g|\over g^2_{\mu-1,\mu-1}}+2\partial_\sigma\bigg)
g^{\sigma\sigma}\partial_\sigma
\left(g_{\mu-1,\mu-1}\sin^2x_{\mu-1}\right)\cr
=&4(n-\mu) -
\bigg[
\big(\sum_{\sigma>\mu-1}\partial_{\mu-1}\ln|g_{\sigma\sigma}|\big)
-2\times 2\cot x_{\mu-1} +2\partial_{\mu-1}\bigg]
\sin 2x_{\mu-1}\cr
&\qquad-\sin^2x_{\mu-1}\sum_{\sigma<\mu-1}
\left(\partial_\sigma\ln{|g|\over g^2_{\mu-1,\mu-1}}+2\partial_\sigma\right)
g^{\sigma\sigma}\partial_\sigma g_{\mu-1,\mu-1}\cr
}
$$
\vfil\eject
$$\eqalign{
=&4(n-\mu)
-\left[2\cot x_{\mu-1}(n-\mu+1)-4\cot x_{\mu-1}\right]
\sin 2x_{\mu-1}-4\cos 2x_{\mu-1}\cr
&\qquad-\sin^2x_{\mu-1}\sum_{\sigma<\mu-1}
\left(\partial_\sigma\ln{|g|\over g^2_{\mu-1,\mu-1}}+2\partial_\sigma\right)
g^{\sigma\sigma}\partial_\sigma g_{\mu-1,\mu-1}
\cr
=&4(n-\mu)-4
\Big[\cos^2x_{\mu-1}(n-\mu+1)-2\cos^2x_{\mu-1}+\cos^2x_{\mu-1}
-\sin^2x_{\mu-1}\Big]\cr
&\qquad-\sin^2x_{\mu-1}
\sum_{\sigma<\mu-1}
\left(\partial_\sigma\ln{|g|\over g^2_{\mu-1,\mu-1}}+2\partial_\sigma\right)
g^{\sigma\sigma}\partial_\sigma g_{\mu-1,\mu-1}\cr
=&4\Big[n-\mu-\cos^2x_{\mu-1}(n+1-\mu) 
+\sin^2x_{\mu-1}+\cos^2x_{\mu-1}\Big]\cr
&\qquad-\sin^2x_{\mu-1}\sum_{\sigma<\mu-1}
\Big(\partial_\sigma\ln{|g|\over g^2_{\mu-1,\mu-1}}
+2\partial_\sigma\Big)
g^{\sigma\sigma}\partial_\sigma g_{\mu-1,\mu-1}\cr
=&4(n+1-\mu)\sin^2x_{\mu-1}
-\sin^2x_{\mu-1}\sum_{\sigma<\mu-1}
\Big(\partial_\sigma\ln{|g|\over g^2_{\mu-1,\mu-1}}
+2\partial_\sigma\Big)g^{\sigma\sigma}\partial_\sigma g_{\mu-1,\mu-1}\cr
=&4R_{\mu\mu}\cr}
$$
\noindent
Next

\noindent
$
4R_{11}=
-\sum_{\sigma>1}
\left[2\partial^2_1\ln|g_{\sigma\sigma}|+
\left(\partial_\mu\ln|g_{\sigma\sigma}|\right)^2\right]
=4\left({1\over \sin^2x_1}-\cot^2x_1\right)
\sum_{\sigma>1}=4(n-1)
$

\noindent
All preceding calculations amount to proving that
$R_{\rho\mu}=(n-1)g_{\rho\mu}$.
  Let $\{ v^\rho_a\}$ be the dual basis of the tangent space where $\rho$
labels a particular basis vector.  
Then 

$R_{ab}=R_{\rho\sigma}v^\rho_av^\sigma_b=(n-1)
g_{\rho\sigma}v^\rho_av^\sigma_b=(n-1)g_{ab}$

\noindent
This fact about $n$-sphere is also provable using the machinery of 
symmetric spaces. See chapter 13 of [2].  

Equations (1) and (2) were derived when attempt was made to 
establish any connection between diagonality of $R_{\nu\sigma}$ and 
that of $g_{\sigma\nu}$. Finally we remark that Dingle [3] has listed 
values of $T^\sigma_\nu$ for the case of 4-dimensional diagonal metric.

\vskip20pt
\noindent
{\bf Acknowledgements}

I would like to thank Professor Jennie 
Traschen for much encouragement and 
discussion and for useful comments on the manuscript.

 \vfill
\eject
\noindent
{\bf References}
\vskip13pt
\noindent
1. Wald,~R.~M. 1984, {\it General Relativity} (Chicago: 
University of Chicago Press).\hfil\break
2. Weinberg,~S. 1972, {\it Gravitation and Cosmology}:{\it ~Principles and 
Applications of The 

General Theory of Relativity} (New York: 
Wiley).\hfil \break
3. Dingle,~H. 1933, ``Values of $T^\nu_\mu$ and the Christoffel Symbols 
for a Line Element of

Considerable Generality,''
 {\it Proc. Nat. Acad. Sci.}, {\bf 19}, 559-563.

\vfill
\end